\documentclass[twocolumn,noshowpacs,preprintnumbers,amsmath,amssymb]{revtex4}
\usepackage{graphicx}
\usepackage{dcolumn}
\usepackage{bm}
\usepackage[english]{babel}
\usepackage[usenames,dvipsnames]{color}
\usepackage[normalem]{ulem}

\begin{document}

\title{Constraining density functional approximations to yield self-interaction free potentials}
\author{Nikitas I. Gidopoulos$^{1,2}$, Nektarios N. Lathiotakis$^2$ \\
\em $^{1}$ISIS, STFC, Rutherford Appleton Laboratory, Didcot, OX11 0QX, United Kingdom\\
$^{2}$Theoretical and Physical Chemistry Institute, National Hellenic Research Foundation, 
Vass. Constantinou 48, GR-11635 Athens, Greece}
\date{\today}

\begin{abstract}
Self-interactions (SIs) are a major problem in density functional approximations 
and the source of serious divergence from experimental results. 
Here, we propose to optimize density functional total energies in terms of 
the effective local potential, under constraints for the effective potential 
that guarantee it is free from SI errors and consequently asymptotically correct.
More specifically, we constrain the Hartree, exchange and correlation 
potential to be the electrostatic potential 
of a non-negative effective repulsive density of $N-1$ electrons. 
In this way, the optimal effective potentials exhibit the correct asymptotic decay, 
resulting in significantly improved one-electron properties.
\end{abstract}


\maketitle

\section{Introduction}

It is already thirty years since Perdew and Zunger in a seminal contribution \cite{PZ} proposed to cure 
the self-interaction (SI) error in density functional approximations (DFAs).
The SI error arises from the incomplete cancellation of the 
self-repulsion of the electron density $\rho$ in the direct Coulomb or 
Hartree energy $U[\rho]$ by the approximate exchange energy functional $E^{\rm DFA}_{\rm X}[\rho]$. 
SI errors manifest in inaccuracies of DFAs \cite{kuemmel_rmp} in many ways, e.g. in the 
calculation of binding energies\cite{stefan}, 
underestimation of activation energy barriers\cite{bar1,bar2}, and in single-particle 
properties like ionization potentials (IPs) \cite{perlev,gu}, electron affinities (EAs) 
(unbound anions) \cite{kier} and band gaps of solids\cite{tera}.

Perdew and Zunger \cite{PZ} proposed a many-body SI correction energy term which, in the limit of 
a single electron, eliminates the SI error exactly. 
Their work initiated the field known as self-interaction corrected 
density functional theory (SIC-DFT). 
Unfortunately, the many-body generalization of the one-electron SI energy 
correction is not unique and to date an unambiguous definition is not available. 
Rigorously, we have a sufficient condition for an approximate exchange 
and correlation (XC) energy density functional $E_{\rm XC}[\rho]$ to be 
$N$-representable \cite{si_yang} and thus free from many-body SI errors. 
A distinction of one- and many-body SI  error was made independently by Ruzsinszky et al\cite{ruzsinszky}.
An important development in this area is the 
appreciation of the underlying relationship between the SI error with the fractional charge error \cite{PZ,perlev,si_yang}.
The approximate treatment of many-body SI errors with SIC-DFT leads traditionally to single-particle equations with 
orbital dependent potentials, i.e. the minimization problem is significantly
more complicated than an iterative diagonalization. 
For solids, SIC-DFT is expressed in terms of maximally localized Wannier states.    
An advantage of removing SI errors is that it improves orbital 
energies\cite{harrison,fujiwara}. These orbital energies, are commonly obtained as the 
eigenvalues of the non-diagonal Lagrange multiplier matrix employed
to enforce orbital orthogonality, although the diagonal values of this matrix have also been 
proposed as appropriate~\cite{vydrov}. 
Despite complications, SIC-DFT has been extensively developed and applied to a large variety of 
systems\cite{strange,hughes,toher,filippeti,vogel,ruzsinszky}.

A constrained minimization of the total energy, so that the occupied Kohn-Sham (KS) orbitals arise as the lowest 
eigenorbitals of a common local multiplicative potential has also 
been developed recently \cite{sicoep}, using the optimized effective potential (OEP) 
method \cite{kuemmel_rmp,sharp,talman,bl}. 


Probably the most serious flaw caused by SI errors lies in the asymptotic behavior of the KS 
potential \cite{perdew1990}.
If the cancellation of SI terms was complete then at infinity, the electron-electron contribution to the
KS potential (Hartree and XC) should be $(N-1)/r$ where $N$ is the number of electrons. 
The physical meaning is obvious, at infinity each electron feels the screening of the nuclear charge 
by the remaining $N-1$ electrons. 
The components of the Hartree potential $v_{\rm H}$
and of the exact XC potential $v_{\rm XC}$ to the asymptotic decay 
are $N/r$, and $-1/r$, respectively. 
However, the asymptotic decay of $v^{\rm DFA}_{\rm XC}$ in typical DFAs, such as the local density approximation (LDA) 
or the generalized gradient approximation (GGA), does not follow a power law 
($-c/r$), but is exponentially fast ($c=0$). 
Consequently an electron at infinity is repelled by an effective charge of $N$ 
rather than $N-1$ electrons.
The incorrect asymptotic behavior has dramatic consequences on one-electron properties like 
the IPs, EAs and the fundamental gaps. It also impairs the optical spectrum through linear response in 
time dependent DFT\cite{tddftbook}. 

In the present work, we propose a quantification of the many-body SI error in the Hartree-exchange and correlation 
potential ($v_{\rm HXC}$) of any DFA (Sec.~\ref{sect:quant}). 
Following our definition, we develop a simple way to address SIs by a constrained OEP minimization of the DFA total energy (Sec.~\ref{sec:method}). 
The constraints on the optimized effective potential remove the effects of SIs from the 
potential and enforce the correct asymptotic behavior. 
Finally, in Sec.~\ref{sec:results}, we present our
implementation and the first numerical results.

\section{Quantifying self-interactions in the effective potential\label{sect:quant}}

Aiming to address in an unambiguous manner the SIs in finite systems   
we decided, rather than dealing with the approximate Hartree 
($U$) and $E_{\rm XC}$ energies which remain unchanged and contaminated with SI errors, 
to focus on the effective local potential. 
The latter (i.e. the HXC potential) is obtained as the functional 
derivative of the HXC energy with respect to the density. 
In KS theory, $v_{\rm HXC} = v_{\rm H} + v_{\rm XC}$ screens the nuclear attraction felt by a KS electron. 
By virtue of Poisson's equation, and following G\"orling \cite{AG_x}, the Laplacian of $v_{\rm HXC}$ defines
the HXC density $\rho_{HXC}$: 
\begin{equation} \label{pois}
\nabla ^2 v_{\rm HXC} ({\bf r}) = -4 \pi \, \rho_{\rm HXC} ({\bf r}) \, . 
\end{equation}
Here, $\rho_{\rm HXC}$ is the density with electrostatic potential $v_{\rm HXC}$.
The integrated charge,
\begin{equation} \label{eq:qxc}
Q_{\rm HXC} = \int d{\bf r}\, \rho_{\rm HXC} ({\bf r})\,,
\end{equation}
allows us to quantify the SI error of the approximate HXC potential.
For example, if $Q_{\rm HXC} = N-1$ then each electron interacts with an effective electrostatic
charge of $N-1$ electrons and this is a necessary condition that the approximation is SI free. 
If $N-1 < Q_{\rm HXC} < N$ there is partial cancellation of SI's and finally if 
$Q_{\rm HXC} = N$, each electron interacts electrostatically with the charge of the other $N-1$ electrons plus
an additional electron that can only be attributed to the same electron itself. 
We say that the corresponding HXC potential exhibits full SI effects. 
In popular approximations, such as LDA or GGA, $Q_{\rm HXC} = N$ as can be deduced 
from the asymptotic behavior of the potential. Thus, according to the present definition 
there is full SI in the HXC potential in these approximations.

Our criterion for SI can be equivalently expressed in terms of the exchange and correlation 
charge, 
$$
Q_{\rm XC} \doteq Q_{\rm HXC} - N \, , 
$$
which must equal $Q_{\rm XC}=-1$ for SI-free approximations. 
This condition was employed by G\"orling \cite{AG_x} to constrain the asymptotic
behavior of a finite-basis set implementation of the exact exchange (EXX) potential.  

The exchange and correlation charge $Q_{\rm XC}$ and the corresponding exchange and correlation density 
\cite{AG_x,parr_x}, $\rho_{\rm XC} ({\bf r}) \doteq \rho_{\rm HXC} ({\bf r}) - \rho ({\bf r})$, 
bear a similar name to the familiar exchange and correlation 
hole of an electron at $\bf r$, $n_{\rm XC}({\bf r},{\bf r}')$\cite{parr_yang}.
The latter is a property directly obtained from the pair correlation function $h ({\bf r},{\bf r}')$, 
\begin{equation} 
n_{\rm XC}({\bf r},{\bf r}') = \rho({\bf r}') \, h ({\bf r},{\bf r}') \, ,
\end{equation}
and satisfies the sum rule $\int d{\bf r}' \, n_{\rm XC}({\bf r}, {\bf r}') = -1 $.

There is no easy and direct relation between the exchange and correlation hole (two-electron property) 
and the exchange and correlation density (one-electron property).
In particular, the approximate LDA XC hole, $n_{XC}^{\rm LDA}({\bf r},{\bf r}')$, satisfies the sum rule
$\int d{\bf r}' \, n_{XC}^{\rm LDA}({\bf r}, {\bf r}') = -1 $, since LDA corresponds to a physical system,
the uniform electron gas, but the satisfaction of this sum rule does not preclude SI errors from the LDA potential, 
$v_{\rm XC}^{\rm LDA}$, except of course when LDA is applied (tautologically) to the uniform electron gas \cite{parr_yang}. 

\section{\label{sec:method} Constraining the optimized effective potential}

In order to correct SIs in DFAs we propose to replace the Hartree, exchange and correlation 
 potential $v^{\rm DFA}_{\rm HXC}$ in the KS equations 
with the effective repulsive potential, $v_{\rm rep}$, obtained from a constrained 
minimization of the DFA total energy,
\begin{equation} \label{ksrep}
\left[
-\frac{\nabla^2}{2} + v_{\rm en} ({\bf r}) + v_{\rm rep}({\bf r})
\right] \phi_i ({\bf r}) = \epsilon_i \, \phi_i ({\bf r}) \, ,
\end{equation}
where $v_{\rm en} ({\bf r})$ is the attractive electron-nuclear potential.

The potential $v_{\rm rep}$ is optimized, in the fashion of the OEP method, 
by requiring that its $N$ lowest orbitals give the density $\rho = \sum_{i=1}^N |\phi_i|^2$ 
that minimizes the DFA total energy.
We constrain the DFA total energy minimization by restricting the potential $v_{\rm rep}$ 
to satisfy two conditions on its effective repulsive density $\rho_{\rm rep}$, i.e. the
density with electrostatic potential $v_{\rm rep}({\bf r})$,
\begin{equation}\label{eq:eff_potential}
v_{\rm rep}({\bf r}) = \int d{\bf r}' \;  \frac{\rho_{\rm rep}({\bf r}')}{|{\bf r}-{\bf r}'|}\,.
\end{equation}
The two conditions read:
\begin{eqnarray}
\label{eq:rep_charge}
Q_{\rm rep} = \int d {\bf r} \; \rho_{\rm rep}({\bf r})& = & N-1 \, , \\
 \rho_{\rm rep}( {\bf r}) & \ge & 0 \, . \label{positive}
\end{eqnarray} 
The system of equations (\ref{eq:rep_charge}) and (\ref{positive}) is equivalent to the 
numerically simpler-to-implement system (\ref{eq:rep_charge}) and (\ref{eq:pos}), with 
\begin{equation}\label{eq:pos}
\int d{\bf r} \; |\rho_{\rm rep}({\bf r})| = N-1 \,.
\end{equation}
We remark that our constrained minimization results in a DFA total energy minimum that 
is in general higher than the global minimum, unless the global minimizing potential happens 
to satisfy the two extra constraints (\ref{eq:rep_charge}) and (\ref{positive}).

As we have discussed,  the normalization of the total effective repulsive 
charge (\ref{eq:rep_charge}) is necessary 
for the absence of SIs from the effective repulsive potential. 
However, Eq. (\ref{eq:rep_charge}) on its own is not sufficient 
to ensure the absence of SI effects in $v_{\rm rep}$, as it would 
be energetically favorable to retain SIs locally near the system, 
plus a compensating negative charge far away from the system to satisfy (\ref{eq:rep_charge}). 
%
Specifically, by imposing this constraint alone, one will still obtain, almost everywhere, 
the global minimum, i.e., $v_{\rm rep} = v_{\rm HXC}^{\rm DFA}$, with 
$\rho_{\rm rep} = \rho_{\rm HXC}^{\rm DFA}$ almost everywhere, integrating to almost $N$ up to a very large 
distance from the system, plus a compensating electronic charge of $-1$ distributed over a large radius, 
very far away from the system, where the addition of the extra negative charge will not cost energetically. 
Since the large radius must be as far away as possible and its position is not well defined, 
the problem strictly has no solution.

With the second (strong) condition, Eq.~(\ref{positive}) or (\ref{eq:pos}), 
such a pathological behavior can be avoided and 
the pair of conditions (\ref{eq:rep_charge}) and (\ref{positive}) becomes sufficient 
 for the absence of SIs from the potential -- although 
probably it is not necessary any more. 
Equation~(\ref{positive}) is an approximate condition but it has an obvious and appealing
physical interpretation: since $\rho_{\rm rep}$ is non-negative and integrates to $N-1$, 
it corresponds to a virtual system of $N-1$ electrons repelling the electron at $\bf r$. 
There is no longer any freedom for our solution 
to collapse to the global minimum solution with a compensating 
negative charge at large distances. Hence the constraint of Eq.~(\ref{positive})
allows for a physical enforcement of the constraint of Eq.~(\ref{eq:rep_charge})
correcting SIs from $v_{\rm rep}$.

We note that if the effective repulsive density is expanded in a small finite basis set 
it is possible that employing (\ref{eq:rep_charge}) alone 
may result in a solution that appears physical.
This is an artifact of the smallness of the basis set, and
by increasing the size of the effective repulsive  density basis set, the pathology of not having 
a sufficient condition will emerge.

In contrast to $v_{\rm HXC}^{\rm DFA} \doteq \delta (U[\rho]+E_{\rm XC}^{\rm DFA}[\rho]) / \delta \rho$, 
the potential $v_{\rm rep}$ is not the functional derivative of $U[\rho]+E_{\rm XC}^{\rm DFA}[\rho]$ and      
it is unknown if $v_{\rm rep}$ is the functional derivative of some 
other functional. This question will be explored in a future work where our method is 
generalized to extended systems. 
Of course, the full KS potential $v_{\rm en} + v_{\rm rep}$  in Eq. (\ref{ksrep}) is 
the functional derivative with respect to the density 
of the non-interacting kinetic energy functional $T_s[\rho]$  at $\rho=\sum_{i=1}^N |\phi_i|^2$, as with
any OEP theory.   

In a related work to correct the asymptotic behavior of the 
effective potential, Wu {\it et al} \cite{Wu} partitioned the effective potential into the Fermi Amaldi potential,
 which  has the correct asymptotic behavior, and they expanded the remainder in a finite localized basis set. 
However, for a large basis set (not complete), the tail of the potential (for moderate distances) will revert 
to that of the unconstrained DFA potential and only for very large distances will the correct asymptotic behavior
be recovered.  

Andrade and Aspuru-Guzik~\cite{xavier}, also aimed at the same problem using an {\it ad hoc} correction of 
the XC density at large distances. 
In our approach the effective repulsive density and potential are obtained directly from the 
minimization procedure.

A feature of our method which contrasts it to SIC-DFT, is that 
the DFA total energy is unchanged and consequently it is invariant under  
unitary transformations of the occupied orbitals. 

Finally, an important test for any theory that corrects SIs 
is that it has the correct one-electron limit. 
In this case, the repulsive potential $v_{\rm rep}$ should vanish for a one-electron system and 
indeed it is straightforward to confirm that constraints 
(\ref{eq:rep_charge}) and (\ref{positive}) give $\rho_{\rm rep} ({\bf r}) = 0$, leading to 
$ v_{\rm rep} ({\bf r}) = 0$, as expected.


Our search for the effective repulsive density and potential 
is performed by expanding them in a basis and then searching for the expansion coefficients,
\begin{equation}\label{eq:xi_basis0}
    \rho_{\rm rep}({\bf r}) = \sum_l v_l\: \chi_l({\bf r}) \, , 
\end{equation}
\begin{equation}\label{eq:xi_basis}
    v_{\rm rep}({\bf r}) = \sum_l v_l\: \xi_l({\bf r}), \;\; \mbox{with}\;\; \xi_l({\bf r}) = \int d{\bf r}'\; 
 \frac{\chi_l({\bf r}')}{|{\bf r}-{\bf r}'|} ,
\end{equation}
where $\chi_l({\bf r})$ is an auxiliary basis set, for example, localized gaussians. 

The minimization of the approximate 
total energy, $E_{\rm DFA}$, with respect to $v_l$ under the conditions (\ref{eq:rep_charge}) 
and (\ref{eq:pos}) leads to the variation equation
\begin{equation}\label{eq:system}
\frac{\partial E_{\rm DFA}}{\partial v_l} = \mu \:X_l + \lambda \:\bar{X}_l\,,
\end{equation}
where $\mu$, $\lambda$ are Lagrange multipliers to satisfy (\ref{eq:rep_charge}), (\ref{eq:pos}) and 
\begin{eqnarray}
X_l &=& \int d{\bf r}\; \chi_l({\bf r}) \, , \\
\bar{X}_l &=& \int d{\bf r} \; \chi_l({\bf r}) \, { |\rho_{\rm rep}({\bf r})| \over \rho_{\rm rep} ({\bf r}) } \, .
\end{eqnarray}

The derivative on the left hand side is obtained, as in the OEP method, through a chain rule:
\begin{equation}
\frac{\partial E_{\rm DFA}}{\partial v_l} = \int \!\!\! \int d{\bf r} \, d{\bf r}' \, 
{\delta E_{\rm DFA} \over \delta v_{\rm rep} ({\bf r}) } \, 
{\delta v_{\rm rep} ({\bf r}) \over \delta \rho_{\rm rep} ({\bf r}') } \, 
{\partial \rho_{\rm rep} ({\bf r}') \over \partial v_l}\,. 
\end{equation}
The functional derivative ${\delta E_{\rm DFA} / \delta v_{\rm rep} ({\bf r}) }$ is obtained analogously to the 
OEP functional derivative, 
\begin{equation}
{\delta E_{\rm DFA} \over \delta v_{\rm rep} ({\bf r}) } = 2
\sum_{i a} {\langle i | v_{\rm HXC}^{\rm DFA}  - v_{\rm rep} | a \rangle \over 
\epsilon_i - \epsilon_a } \,  \phi_i ({\bf r}) \phi_a ({\bf r})\,,
\end{equation}
where $v_{\rm HXC}^{\rm DFA} \doteq \delta (U[\rho]+E_{\rm XC}^{\rm DFA}[\rho]) / \delta \rho $, 
$i$ runs over occupied, $a$ over unoccupied eigenorbitals of $v_{\rm rep}$, 
with $\epsilon_i$ and  $\epsilon_a$ the corresponding eigen-energies. 
We also have that 
\begin{equation}
{\delta v_{\rm rep} ({\bf r}) \over \delta \rho_{\rm rep} ({\bf r}') } = {1 \over |{\bf r} - {\bf r}' | } \,.
\end{equation}
We finally obtain
\begin{eqnarray} 
\frac{\delta E_{\rm DFA}}{\delta v_l} &=& 2 \sum_{ia} \frac{ v^{\rm HXC}_{ia} - v^{\rm rep}_{ia} }
{\epsilon_i -\epsilon_a}\: S^{(l)}_{ia} \, , \label{eq:funcder} \\ 
 S^{(l)}_{ia} &=& \int d{\bf r}\: \phi_a({\bf r}) \: \phi_i({\bf r}) \: \xi_l({\bf r}) \, .
\end{eqnarray}
Here, $v^{\rm rep}_{ia}$ and $v^{\rm HXC}_{ia}$ are the matrix elements
of the potentials $v_{\rm rep}$ and 
$v_{\rm HXC}^{\rm DFA} $, 
respectively. 
Eqs.~(\ref{eq:system}) and (\ref{eq:funcder}) define a non-linear system of equations with respect to
$v_l$. This system can be solved, using an iterative scheme of two steps.
In the first step, a linear system is solved by keeping the quantities 
$\phi_i$, $\phi_a$, $\epsilon_i$, $\epsilon_a$, $v^{\rm HXC}_{ia}$, and $S^{(l)}_{ia}$ 
frozen. In the second step, a single-electron Hamiltonian problem is solved with the 
potential obtained in the previous step and the quantities $\phi_i$, $\phi_a$, $\epsilon_i$, 
$\epsilon_a$, $v^{\rm HXC}_{ia}$, and $S^{(l)}_{ia}$ are updated. Our numerical implementation
proved that this scheme is very efficient and usually only a few iterations are required to converge 
using a mixing scheme similar to Kohn-Sham iterative procedure.
The potential obtained at the first step, i.e. when orbitals are frozen to the Kohn Sham orbitals
is already a very good approximation to the local potential.
The linear system that needs to be solved in each iteration has the form 
\begin{equation}\label{eq:linsys}
\sum_l A_{kl} v_l = b_k+\mu \:X_k + \lambda \:\bar{X}_k\,, 
\end{equation}
with
\begin{equation}\label{eq:linsys1}
A_{kl} = \sum_{ia} \frac{S^{(k)}_{ia}\:S^{(l)}_{ai}}{\epsilon_i -\epsilon_a}\,, \mbox{\ and\ \ } 
b_{kl} = \sum_{ia} \frac{S^{(k)}_{ia}\:v^{\rm HXC}_{ia}}{\epsilon_i -\epsilon_a}\,.
\end{equation}
The Lagrange multipliers $\mu$, $\lambda$ are given by the solution of a simple 2$\times$2 linear system
obtained from Eq.~(\ref{eq:linsys}) by multiplying both sides by the inverse of $A$, then by $X_k$ 
(or $\bar{X}_k$) and summing over $k$, and using Eqs.~(\ref{eq:rep_charge} and \ref{eq:pos}).


Eqs.~(\ref{eq:linsys}) and (\ref{eq:linsys1}) constitute a simple modification of the 
usual OEP equations. 
The solution of Eq.~(\ref{eq:linsys}) is complicated because often the matrix $A$ 
is singular. This problem is well known in the OEP method with finite basis sets 
\cite{davidson,hirata,hess,filatov1,glushkov2009}, 
and the solution involves the inversion of $A$ in the space of its
non-singular eigenvectors, usually with a singular value decomposition (SVD).
However, even after the SVD, and depending on the particular basis sets, 
the resulting effective potential may look unphysical.
In Ref. \onlinecite{our}, we argue that 
in addition to the known technical problem of inversion of $A$ 
lies an unexpected discontinuity of the optimal potential, 
when the orbital basis set is truncated with a finite basis.
In the present work, the effect of this discontinuity is 
reduced significantly by the restriction of the admissible 
potentials to satisfy conditions (\ref{eq:rep_charge}) and especially (\ref{eq:pos}).

\section{\label{sec:results}Numerical Applications}

Our numerical implementation is based on a Gaussian basis set expansion for the
orbitals as well as for the effective potentials or for the effective repulsive density. 
XC functionals are provided by the Libxc library\cite{libxc}.

\subsection{Constraining the EXX potential.}

\begin{figure*} 
\begin{tabular}{cc}
\includegraphics [width=0.9\columnwidth]{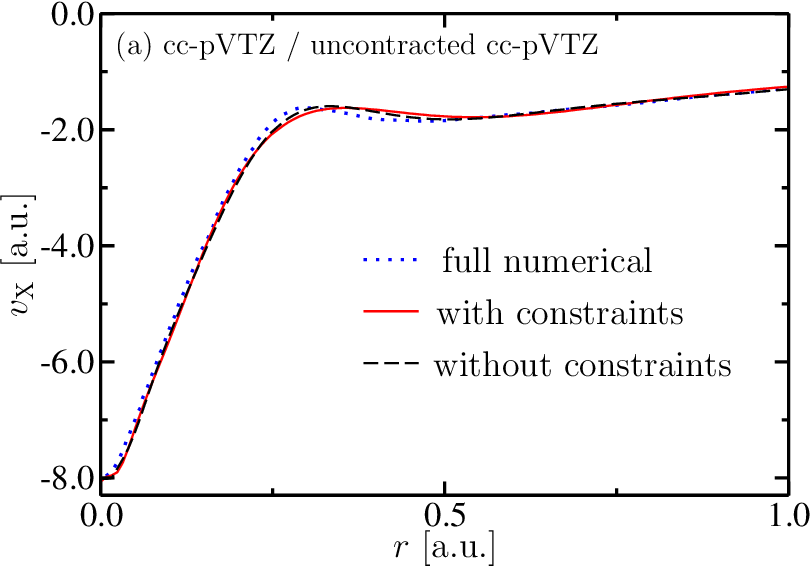} \ & \ 
\includegraphics [width=0.9\columnwidth]{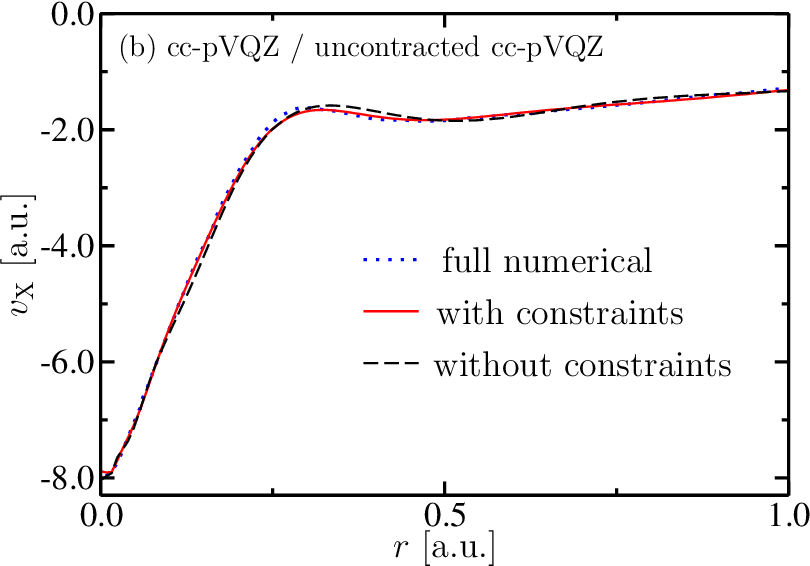} \\
\end{tabular}
\caption{
Exchange part, $v_{\rm X}$, of the effective repulsive potential for the 
Ne atom using Eqs.~(\ref{eq:rep_charge}) and (\ref{positive}) (constrained potential) 
compared to the unconstrained finite basis EXX potential and the full numerical result or Ref.~\onlinecite{hess}. 
For the constrained and unconstrained EXX potentials, we used our amended finite basis OEP equations in Ref.~\onlinecite{our}.
\label{fig:exx}}
\end{figure*}

A rigorous test is to apply our constrained OEP method to a functional that is free from
SIs such as the exact exchange functional and compare the 
potentials obtained with and without the constraints 
of Eqs.~(\ref{eq:rep_charge}) and (\ref{positive}). 
This comparison is shown in Fig. \ref{fig:exx} for the Ne atom.
To obtain the unconstrained potentials in the two plots, the Hartree and exchange potential 
(i.e. not the effective repulsive density) was expanded directly 
in the uncontracted cc-PVTZ and uncontracted cc-PVQZ basis sets respectively.
For the constrained case, the effective repulsive densities were expanded 
in the same basis sets. The orbitals were expanded in the cc-PVTZ and cc-PVQZ basis sets.
  
In the unconstrained minimization case, to obtain reasonable EXX potentials with the finite basis sets, 
we employed the amended finite basis OEP Eq.~(40) in Ref.~\onlinecite{our} 
which contains an extra term (discontinuity correction, with $\lambda=10^{-3}$) 
that restores continuity of the potential. The unconstrained potentials contain 
an arbitrary constant and were shifted so that the energy of the
highest occupied molecular orbital (HOMO) equals that of Hartree
Fock (HF). As can be seen in Fig.~\ref{fig:exx}, the unconstrained finite basis EXX potentials 
are very close to the full numerical EXX potential from Ref.~\onlinecite{hess}.

For the constrained minimization case, the discontinuity of the OEP with finite basis sets 
is reduced because the variational freedom of the admissible potentials is 
restricted by the two constraints. 
Then, inclusion of the 
discontinuity correction of Ref.~\onlinecite{our} is not
necessary to obtain smooth potentials. 
However, subtle features of the EXX potential, such as the shell structure do not appear 
without the more complete description of OEP including the discontinuity correction. Thus,
for a meaningful comparison between the two results we show the constrained EXX potentials 
employing the discontinuity correction with $\lambda=10^{-3}$.  The two constrained potentials 
are almost on top of the unconstrained potentials and the full numerical solution
as seen in Fig.~\ref{fig:exx}.
In addition, the IP of Ne, given as the minus of the energy of the HOMO of the 
constrained potential, is almost identical to that of HF theory ($\sim$0.2~eV lower) with 
the non-local exchange potential. This 
should be contrasted to the dramatic effect of our constraints on the IPs in the case of 
LDA, i.e. a DFA with full SIs, as we will see in Sec.~\ref{sec:lda}.

To summarize this test, our results indicate that 
the constraints introduced here do not have a significant effect on the potential 
for theories that are free from SIs.
In Sec~\ref{sec:lda}, we apply our methodology to LDA as an example of a DFA that is contaminated with SI errors.
As we shall see in Fig.~\ref{fig:Ne_pot}, the extra constraints modify the LDA KS potential substantially.
  
\subsection{Constraining the LDA potential\label{sec:lda}}

To illustrate our approach we chose the LDA functional\cite{slater_vwn}
and we refer to the combined method as constrained LDA (CLDA).
The LDA-KS potential misses the shell structure present in EXX even after the implementation of our SI 
correction. For CLDA, the slight improvement in accuracy afforded by the inclusion of the 
discontinuity correction of Ref.~\onlinecite{our} was not sufficient to warrant its use. 
We found that a simple SVD and inversion of the matrix $A$ in Eq.~(\ref{eq:linsys}) gave adequate accuracy. 


To perform the SVD of $A$, we divide the space spanned by the
eigenvectors of $A$ in the null space and the rest, 
using a small parameter $\theta$ as a cut-off for the null eigenvalues. 
For the systems and basis sets we considered, $\theta = 10^{-6}$ has proven a reasonable choice.

\begin{figure} 
\vspace{0.3cm}
\includegraphics [width=0.45\textwidth]{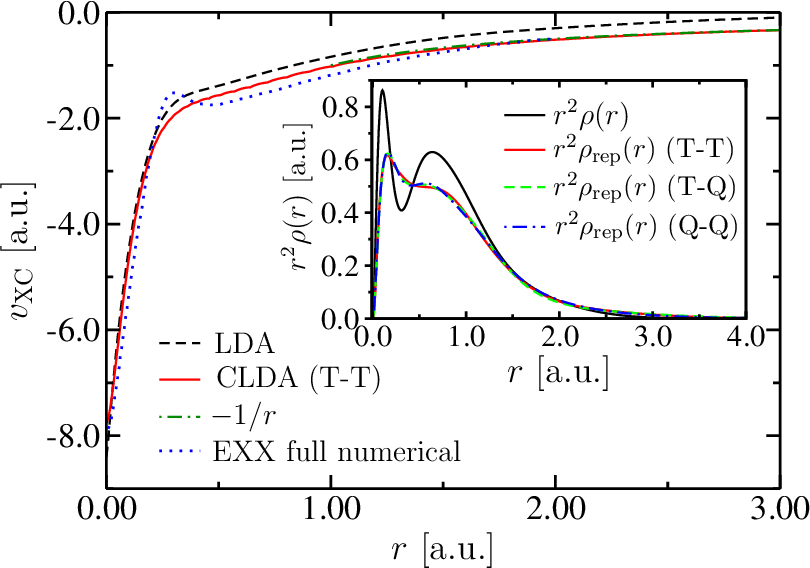}
\caption{ 
The XC part, $v_{\rm XC}$, of the CLDA-effective and KS-LDA potentials as well as the exchange part of
 the full numerical EXX potential (from Ref. \cite{hess}) for the Ne atom. 
$-1/r$ is also shown. In the inset, the effective repulsive charge density is shown.
The notation (X-Y)
stands for cc-pVXZ and uncontracted cc-pVYZ for the expansion of the orbitals and the 
effective repulsive charge density respectively.
\label{fig:Ne_pot}}
\end{figure}

\begin{table}
\begin{tabular}{l|c|c|c|c|c|c}
\hline
 & Basis & $Q_n$ & $\Delta E$ & IP(LDA) & IP(CLDA) & Exp \\
\hline\hline
He        &  T-Q  & 0 &$1.5\cdot 10^{-3}$ & 15.46 & 23.14&  24.6   \\
Be        &  T-T  & $3.0\cdot 10^{-4}$ &$1.1\cdot 10^{-4}$ & 5.59 & 8.62 &  9.32   \\
Ne        &  T-T  & 0 &$2.7\cdot 10^{-5}$ & 13.16&  18.94 &  21.6   \\
H$_2$O    &  T-T  & $6.0\cdot 10^{-5}$ &$1.1\cdot 10^{-5}$ & 6.96&   11.24 & 12.8   \\
NH$_3$    &  T-T  & $6.0\cdot 10^{-5}$ &$8.2\cdot 10^{-6}$ & 6.00&   9.81 &  10.8  \\
CH$_4$    &  D-D  & $1.5\cdot 10^{-3}$ &$2.7\cdot 10^{-4}$ & 9.28&  12.52  &  14.4  \\
C$_2$H$_2$&  D-D  & $1.9\cdot 10^{-4}$ &$4.1\cdot 10^{-5}$ & 7.02&   10.63 &  11.5  \\
C$_2$H$_4$&  D-D  & $3.9\cdot 10^{-3}$ &$1.1\cdot 10^{-3}$ & 6.67&   9.57  &  10.7  \\
CO        &  D-D  & $2.0\cdot 10^{-5}$ &$3.6\cdot 10^{-4}$ & 8.75&   12.73 &  14.1 \\
NaCl      &  D-D  & $1.2\cdot 10^{-2}$  &$6.8\cdot 10^{-4}$ & 5.13&   7.87  & 8.93 \\
\hline\hline
F$^-$&T\footnotemark[1]-T & $1.0\cdot 10^{-4}$ & $2.7\cdot 10^{-5}$ & $E_{\rm H}>0$  & 2.23 & 3.34 \\
Cl$^-$&T\footnotemark[1]-T & $1.0\cdot 10^{-5}$ & $1.6\cdot 10^{-4}$ & $E_{\rm H}>0$  & 2.61 & 3.61\\
OH$^-$&T\footnotemark[1]-T & $4.0\cdot 10^{-5}$ & $1.4\cdot 10^{-4}$ & $E_{\rm H}>0$  & 0.99 & 1.83\\
NH$_2^-$&T\footnotemark[1]-T & $4.0\cdot 10^{-5}$ & $8.2\cdot 10^{-4}$ & $E_{\rm H}>0$  & 0.18 &  0.77 \\
CN$^-$&T\footnotemark[1]-T       & $1.0\cdot 10^{-5}$ & $1.1\cdot 10^{-4}$ & 0.13 & 2.87 &  3.77 \\
\hline
\end{tabular}
\footnotetext[1]{For negative ions, aug-cc-pVXZ basis sets were used for the orbital expansion.}
\caption{
\label{table:results}
\label{tab:IP}
The total energy difference $\Delta E$ of CLDA from plain LDA the total negative effective repulsive  charge $Q_n$ (in e)
and the IPs for selected atoms,  molecules (top) and negative ions (bottom). 
IPs and EAs were calculated as the negative of the one-electron energies corresponding to the HOMO of the neutral system and the negative ion respectively. 
Basis set notation is explained 
in the caption of Fig.~\ref{fig:Ne_pot}. For the neutral systems we compare with experimental 
values of the IP, while for the negative ions with experimental values of the EA 
of the corresponding neutral system. All energies are in eV.
}
\end{table}

\begin{figure} 
\vspace{0.3cm}
\begin{tabular}{c}
\includegraphics [width=0.45\textwidth]{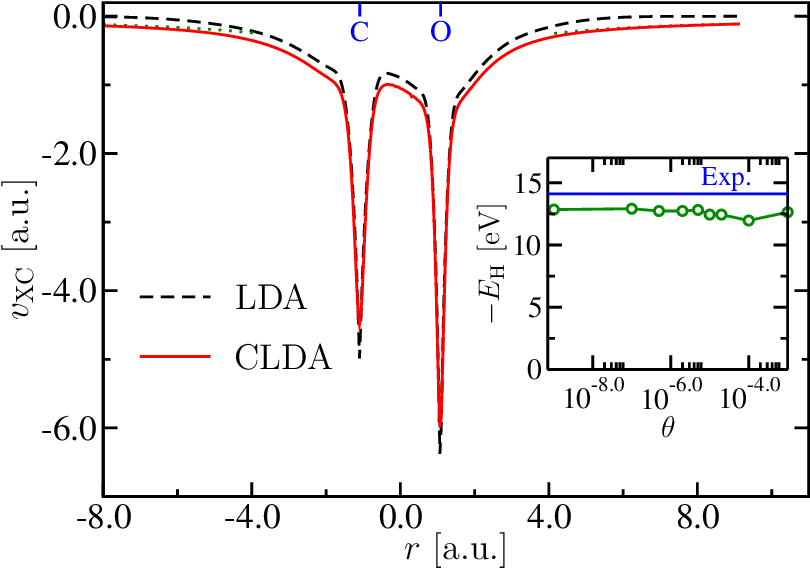} \\ 
\end{tabular}
\caption{
 The XC part, $v_{\rm XC}$, of the effective potential for CO molecule along the molecular axis. Green dotted lines indicate the correct $\pm 1/r$ asymptotic behavior with $r$ measured from the molecular center.  In the inset, the $E_{\rm H}$ as a function of the SVD parameter $\theta$ is shown with the horizontal blue line indicating the experimental value of the IP.
\label{fig:CO_C2H2_pot}}
\end{figure}

In Fig.~\ref{fig:Ne_pot}, we show the LDA and CLDA potentials for the Ne 
atom using finite basis sets as well as the effective repulsive density. 
Evidently, the effective potential obtained with CLDA has the correct asymptotic behavior. 
Also, the effective repulsive density is converged with respect to the different 
basis sets. 
Potentials with the correct asymptotics are also obtained for larger systems like
CO molecule as shown in Fig.~\ref{fig:CO_C2H2_pot}.

In the top of Table~\ref{tab:IP}, we show the IPs calculated with CLDA and with LDA,
 as the negative of the one-electron energy of the HOMO, 
$E_{\rm H}$, for various atomic and molecular systems. 
In the inset of Fig.~\ref{fig:CO_C2H2_pot} we show that the IP value for CO (calculated in that 
way) is essentially independent of $\theta$.

The IPs from CLDA are on average roughly 10\% underestimated. This
divergence should be contrasted to the dramatic 40\% errors of plain LDA. Given the severe 
underestimation of IPs and the fundamental gaps of solids by LDA and GGA, our approach
offers a significant qualitative improvement. 
Contrary to LDA, negative ions are predicted to be bound by CLDA as shown in the 
bottom of Table~\ref{tab:IP}. 
Even though EAs of neutral systems (predicted as the negative of the one-electron energy of the
HOMO of the corresponding negative ions), are underestimated compared 
to experiment by about 40\%, it is nevertheless encouraging 
that qualitatively correct EAs can be obtained with CLDA.

The differences of the total energies obtained with CLDA from those obtained with LDA are also shown 
in Table~\ref{tab:IP}. These differences are very small for all systems, i.e., the additional constraint 
on the potential does not change the total energy. 
The total negative effective repulsive charge, $Q_n$,  is shown in Table~\ref{tab:IP} 
as a measure of how well the positivity condition is fulfilled by the optimal potential. 
Although we did not manage to eliminate it completely, $Q_n$ is very 
small compared to the total effective repulsive charge and does not affect 
the quality of our effective potential. 

As already mentioned, for CLDA we could have used the more sophisticated discontinuity 
correction of Ref.~\onlinecite{our} instead of the SVD. 
To assess if this is required, we applied the modified OEP equation (40) 
of Ref.~\onlinecite{our} to the Ne atom, adopting the same basis sets, and $\lambda=10^{-3}$ as in the case of EXX. 
The obtained 
potential is almost on top of the CLDA potential shown in Fig.~\ref{fig:Ne_pot} while the IP, at 
19.6~eV, is only slightly improved compared to the result in Table~\ref{tab:IP}. 
Thus, the choice of a simple SVD seemed to us reasonable for this first demonstration of our approach. 

\section{Discussion}

Until now, the main errors stemming from SIs were not in the total energy but resulted from 
deficiencies of the local potential. 
The main advantage of splitting the XC energy $E_{\rm XC}=E_{\rm X}+E_{\rm C}$ 
and treating the exchange exactly is the cancellation of SIs 
and the quality of the KS potential, however at a computational cost compared with LDA/GGA, 
and an ensuing complicated description of correlation through non-local orbital functionals
\cite{pra}. 
In fact an appropriate non-local $E_{\rm C}[\rho]$ of cost no-higher than EXX is not available yet.

Attempting to address the SI problem in DFAs for finite systems, we noted the ambiguity 
in the quantification of the SI error in the Hartree and XC energies.
Still, it was possible to quantify the SI error in the KS potential in a way that is 
unambiguous and independent of the error in the energy.
Following our definition of the SI error in the KS potential, we proposed 
two constraints for the repulsive part of the KS potential: the corresponding effective 
repulsive density must integrate to $N-1$ electrons and it must 
be a non-negative function everywhere. 
The constraint for the norm of the effective repulsive density is 
a necessary property satisfied, for example, by the exact KS and by the EXX potentials 
but it is not a sufficient condition on its own. 
The norm together with the positivity constraint are 
sufficient conditions for the absence of SIs from the potential, 
although the positivity constraint is probably 
too strong and not strictly satisfied by 
the exact KS potential and EXX. 
Since our treatment is still approximate, the ambiguity in dealing with SIs remains. 

Nevertheless, imposing these constraints constitutes a powerful method because the 
constraints restrict considerably the variational freedom of the effective potential 
while at the same time allowing for the accurate description of EXX. 
When our constraints are applied to the EXX energy functional 
they produce potentials that are very close to the EXX potential obtained without any constraint. 
In particular, the constrained EXX potential in Fig.~\ref{fig:exx} preserves the 
shell structure of the numerical potential, by showing the corresponding bump at almost the
same position. This excellent agreement, given the finite nature of the basis sets, shows that
the constraints introduced here do not wash away the atomic shell structure
from the optimal potential which is a subtle but essential feature. 
The fact that this structure is absent from CLDA should be 
traced back to the approximate nature of the LDA functional.

Finally, our constraints keep the scaling of computational cost at the level of the corresponding DFA 
and eliminate the effects of SIs from the potential (where it matters) 
with a minimal increase of the total energy.
For LDA, the constrained KS potentials have the correct asymptotic behavior 
and give significantly improved IPs over the unconstrained LDA results.
At the same time, the description of XC has been kept together 
which has advantages, for example it exploits the cancellation of errors in $E_{\rm XC}$ 
and provides an improved description of electron-pair bonds \cite{bae}.


It is accepted that the many-body SI error leads to spurious fractionally-charged atoms in the 
dissociation of a hetero-nuclear molecule. 
To prevent this spurious effect, the HXC potential must develop a barrier or a step 
between the atoms, as shown by Perdew, in Ref. \onlinecite{jpp}.  
These steps or barriers are related with the shell-structure in atoms; as we have seen in Figs. 
\ref{fig:exx}, \ref{fig:Ne_pot}, the latter is reproduced 
by our constraints in the case of EXX but not in LDA. 
Further investigation is necessary to examine whether our method leads to fractionally 
charged atoms in the dissociation of molecules. 
It is almost certain that this issue depends on the overall quality of the 
approximate energy functional, and does not rely merely on the absence of SIs from the potential.

Other interesting questions regarding our approach are its size consistency and how it applies to extended 
systems where $N$ and $N-1$ are the same. 
All these questions and challenges are related with the energetic cost to localize the 
XC density. In order to address them, our method needs to be extended to include 
a (SI) correction energy term based on the correction of the effective potential. 
This is the focus of work which is in progress.

\begin{acknowledgments}

N.I.G. thanks the TPCI, NHRF for hospitality during an extended visit when part of this work was done.

\end{acknowledgments}


\begin{thebibliography}{99}
\bibitem{PZ} 
J. P. Perdew, A. Zunger, Phys. Rev. B {\bf 23}, 5048 (1981).

\bibitem{kuemmel_rmp} 
S. K\"ummel, L. Kronik, Rev. Mod. Phys. {\bf 80}, 3 (2008).

\bibitem{stefan} 
S. Kurth, J. P. Perdew, P. Blaha, Int. J. Quant. Chem. {\bf 75}, 889 (1999).

\bibitem{bar1} 
K. D. Dobbs, and D. A. Dixon, J. Chem. Phys. {\bf 98}, 12584 ͑(1994͒)

\bibitem{bar2} 
T. N.  Truong and W. Duncan, J. Chem. Phys. {\bf 101}, 7408 ͑(1994͒).

\bibitem{perlev} 
J. P. Perdew, R. G. Parr, M. Levy, J. L. Balduz, Jr., Phys. Rev. Lett. {\bf 49}, 1691 (1982).

\bibitem{gu} 
S. Goedecker, C. J. Umrigar, Phys. Rev. A {\bf 55}, 1765 (1997). 

\bibitem{kier}
M.C. Kim, E. Sim, K. Burke, J. Chem. Phys. {\bf 134}, 171103, (2011).

\bibitem{tera} 
K. Terakura, T. Oguchi, A. R. Williams, and J. K\"ubler, Phys. Rev. B {\bf 30}, 4734 (1984).

\bibitem{si_yang}
P. Mori-S{\' a}nchez, A.J. Cohen, W. Yang, J. Chem. Phys., {\bf 125}, 201102 (2006).

\bibitem{ruzsinszky} 
A. Ruzsinszky, J. P. Perdew, G. I. Csonka, G. E. Scuseria, and O. A. Vydrov, 
Phys. Rev. A {\bf 77}, 060502(R) (2008).

\bibitem{harrison} 
J. G. Harrison, R. A. Heaton, and C. C. Lin, J. Phys. B, {\bf 16}, 2079 (1983).

\bibitem{fujiwara}
T. Fujiwara, M. Arai, Y. Ishii, in {\it Strong Coulomb correlations in electronic structure 
calculations}, V. I. Anisimov, ed. Gordon and Breach (2000), p. 167.

\bibitem{vydrov} 
O. A. Vydrov, G. E. Scuseria, and J. P. Perdew, J. Chem. Phys. {\bf 126} 154109 (2007).

\bibitem{strange} 
P. Strange, A. Svane, W.M. Temmerman, Z.Szotek and H. Winter, Nature {\bf 399} 756 (1999).

\bibitem{hughes} 
I. D. Hughes, et al,
Nature {\bf 446} 650 (2007)

\bibitem{toher} 
C. Toher, and S. Sanvito, Phys. Rev. Lett. {\bf 99}, 056801 (2007).

\bibitem{filippeti} 
A. Filippetti and N. Spaldin, Phys. Rev. B67, 125109 (2003).

\bibitem{vogel} 
D. Vogel, P. Kr\"uger, J. Pollman, Phys. Rev. B {\bf 54}, 5495 (1996).

\bibitem{sicoep}
T. K\"orzd\"orfer, M. Mundt, and S. K\"ummel, Phys. Rev. Lett. {\bf 100}, 133004 (2008)

\bibitem{sharp}
R.T. Sharp, G.K. Horton, Phys. Rev. {\bf 90}, 317, (1953).

\bibitem{talman}
J.D. Talman, W.F. Shadwick, Phys. Rev. A {\bf 14}, 36 (1976).

\bibitem{bl}
F.A. Bulat, M. Levy, Phys. Rev. A {\bf 80}, 052510, (2009).

\bibitem{perdew1990} 
J. P. Perdew, Adv. Quantum Chem. {\bf 21}, 113 (1990).

\bibitem{tddftbook} M.A.L. Marques, C.A. Ullrich, F. Nogueira, A. Rubio, K. Burke, and E.K.U. Gross (eds.), {\it Time-Dependent Density Functional Theory}, Lect. Notes Phys. 706, (Springer, Berlin Heidelberg 2006).

\bibitem{parr_yang}
R. G. Parr, W. Yang, {\it Density-Functional Theory of Atoms and Molecules}, Oxford University Press, Oxford (1989).

\bibitem{AG_x}
A. G\"orling, Phys. Rev. Lett. {\bf 83}, 5459, (1999).

\bibitem{parr_x}
S. Liu, P.W. Ayers, R.G. Parr, J. Chem. Phys., {\bf 111}, 6197 (1999).

\bibitem{Wu} Q. Wu, P. W. Ayers, and W. Yang, J. Chem. Phys. {\bf 119}, 2978 (2003).

\bibitem{xavier} X. Andrade and A. Aspuru-Guzik, Phys. Rev. Lett. {\bf 107}, 183002 (2011).

\bibitem{davidson}
V. N. Staroverov, G. E. Scuseria, E. R. Davidson, J. Chem. Phys. {\bf 124}, 141103, (2006).

\bibitem{hirata}
S. Hirata et al, 
J. Chem. Phys. {\bf 115}, 1635, (2001).


\bibitem{hess}
A. He{\ss}elmann, A.W. G\"otz, F. Della Sala, A. G\"orling, J. Chem. Phys. {\bf 127}, 054102 (2007).


\bibitem{filatov1}
C. Kollmar, M. Filatov, J. Chem. Phys. {\bf 127}, 114104, (2007).

\bibitem{glushkov2009}
V.N. Glushkov, S.I. Fesenko, H.M. Polatoglou, Theor. Chem. Acc. {\bf 124}, 365, (2009)


\bibitem{our}
N. I. Gidopoulos, N. N. Lathiotakis, Phys. Rev. A {\bf 85}, 052508 (2012). 

\bibitem{slater_vwn}
J. C. Slater, The Self-Consistent Field for Molecules and Solids: 
Quantum Theory of Molecules and Solids, vol. 4, 
McGraw-Hill, New York (1974); S.H. Vosko, L. Wilk and M. Nusair. Can. J. Phys. 
{\bf 58}, 1200, (1980).

\bibitem{libxc} 
M.~A.~L. Marques, M.~J.~T. Oliveira, Tobias Burnus, e-print: arXiv:1203.1739;\\ 
http://www.tddft.org/programs/octopus/wiki/index.php/Libxc

\bibitem{bae}
E.J. Baerends, O.V. Gritsenko, J. Chem. Phys., {\bf 123}, 062202 (2005). 

\bibitem{pra}
N.I. Gidopoulos, Phys Rev A, {\bf 83}, 040502(R) (2011).

\bibitem{jpp}
J.~P. Perdew, in {\it Density Functional Methods}, eds. R.M. Dreizler and J. da Providencia, page 265  (Plenum 1985). 



\end{thebibliography}
\end{document}